\documentclass{moriond}

\usepackage{amsmath,amssymb,color,slashed,graphicx}
\def\be{\begin{equation}}
\def\ee{\end{equation}}
\def\bea{\begin{eqnarray}}
\def\eea{\end{eqnarray}}

\newcommand{\Photo} 

\begin{document}
\vspace*{0.3cm}
\title{PROBING RADIATIVE NEUTRINO MASS MODELS AT THE LHC VIA
TRILEPTON EVENTS}
\author{DOUNIA CHERIGUI, CHAHRAZED GUELLA}
\address{Department of Physics, University of Sciences and Technology of Oran, BP
1505, Oran, Algeria}
\author{AMINE AHRICHE}
\address{Department of Physics, University of Jijel, PB 98 Ouled Aissa, DZ-18000 Jijel, Algeria}
\author{SALAH NASRI}
\address{Department of physics, United Arab Emirates University, Al-Ain, UAE.}
\maketitle\abstract
{Trilepton event represents one of the probes of the new physics at high energy colliders. In this talk, we consider the search for processes with final states       $\ell_{\alpha}^{\pm }\ell_{\beta}^{\pm}\ell_{\gamma}^{\mp}+\not\slashed E_{T}$ where
 ${\alpha}$, ${\beta}$, ${\gamma}$= $e,\mu,\tau$, via the production of singlet charged scalar $S^{\pm}$ which arise in a class of radiative neutrino mass models. We discuss the opposite sign same flavor leptons signal, as well as the background free channel in view to get a significant excess at $\sqrt{s}$= 8 TeV and $\sqrt{s}$ = 14 TeV at the hadron collider LHC.}

\section{Introduction}

Accommodating the data from neutrino oscillation experiments required an extension of the standard model of particle with extra degrees of freedom. One of the mechanisms that generate tiny masses for neutrinos, $m_{\nu}$, invoke new  physics at the TeV scale where $m_{\nu}$ vanishes at the tree level but gets generated at higher loop level~\cite{three-loop}. Here, using trilepton events we investigate the
possibility of probing a class of models motivated by neutrino mass
at the LHC. This class of models contains a singlet charged scalar
($S^{\pm}$) that decays to charged lepton and neutrino via
$f_{\alpha\beta}$ Yukawa couplings, inducing lepton flavor violating
(LFV) processes, whose strength is a subject of severe experimental
constraints.

\section{Model Framework \& Space Parameter}

\begin{figure}[h]
\begin{center}
\includegraphics[width=0.43\textwidth]{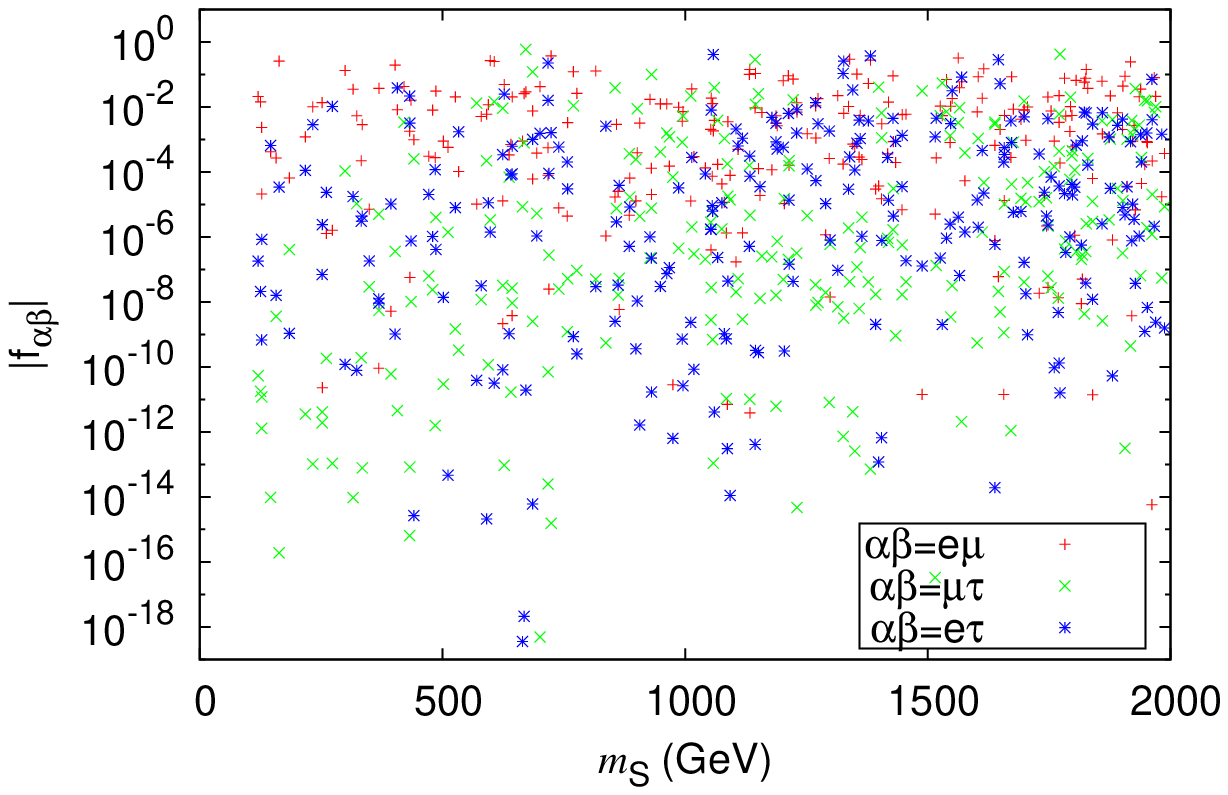}~\includegraphics[width=0.43%
\textwidth]{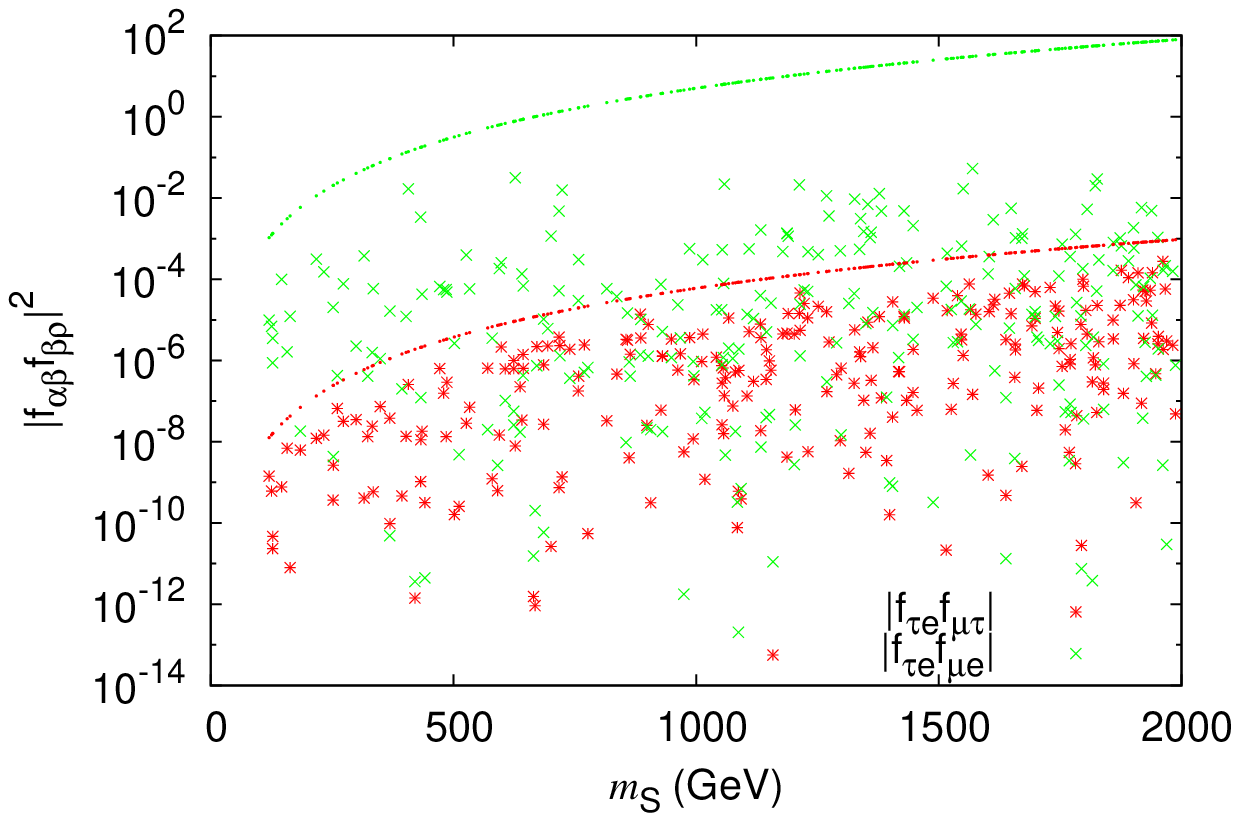}
\caption{Different couplings $f$'s combinations (as absolute values) versus $%
m_{S}$, the experimental bounds $\mu\rightarrow e\gamma$\ and $\tau
\rightarrow\mu\gamma$ are represented by dashed lines.}\label{fab}
\end{center}
\end{figure}

In this work, we consider a class of models that contain the following term
in the Lagrangian%
\begin{equation}
\mathcal{L}\supset f_{\alpha \beta }L_{\alpha }^{T}C\epsilon
L_{\beta }S^{+}-m_{S}^{2}S^{+}S^{-}+\mathrm{h.c.}, \label{eq:LL}
\end{equation}%
The interactions above induce LFV effects via
processes such as $\mu \rightarrow e\gamma $ and $\tau \rightarrow \mu
\gamma $, with the following branching fractions, these two branching ratios should not exceed the
upper bounds $\mathcal{B}\left( \mu \rightarrow e+\gamma \right) <5.7\times
10^{-13}$ ~\cite{Adam:2013mnn} and $\mathcal{B}\left( \tau \rightarrow \mu
+\gamma \right) <4.8\times 10^{-8}$~\cite{PDG}. Moreover, a new contribution to the muon's anomalous magnetic moment is induced at 1-loop.  Figure.~\ref{fab} shows the allowed space parameter for the charged scalar mass range $100~\mathrm{GeV}<m_{S}<2~\mathrm{TeV}$, while scanning over the couplings $f$'s with the LFV  constraints being satisfied.

\section{Current Constraints on Trilepton Signal at the LHC}

\begin{figure}[h]
\begin{center}
\includegraphics[width=0.9\textwidth]{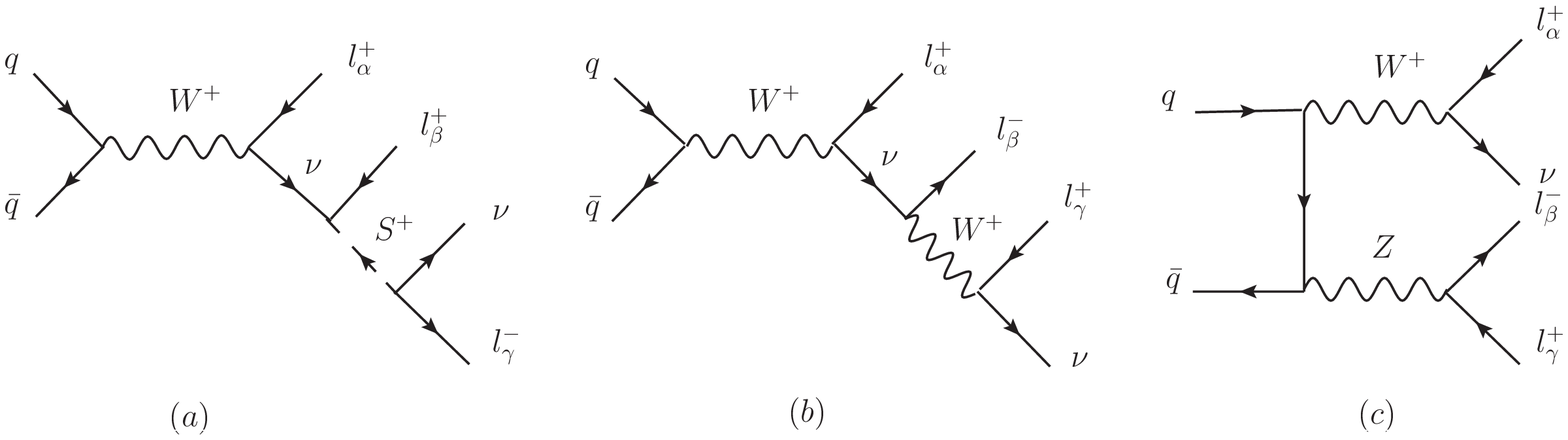}
\end{center}
\caption{Diagrams corresponding to the trilepton signal (a) and SM
background (b,c).}
\label{fig:tri-lepton}
\end{figure}

The charged charged $S^\pm$ can be produced at the LHC through the
processes associated with different sign different flavor charged
leptons at the parton level as shown in
Figure.~\ref{fig:tri-lepton}, including gauge bosons
$W^+Z/W^+\gamma^*$ production as standard model contribution. The subsequent decay of $S^{\pm}$ results in trilepton final states and a missing energy. In our analysis we consider just $\ell=e,\mu$, and use CalcHEP to
generate both searched signal and background events. 

\begin{figure}[h]
\begin{center}
\includegraphics[width=0.43\textwidth]{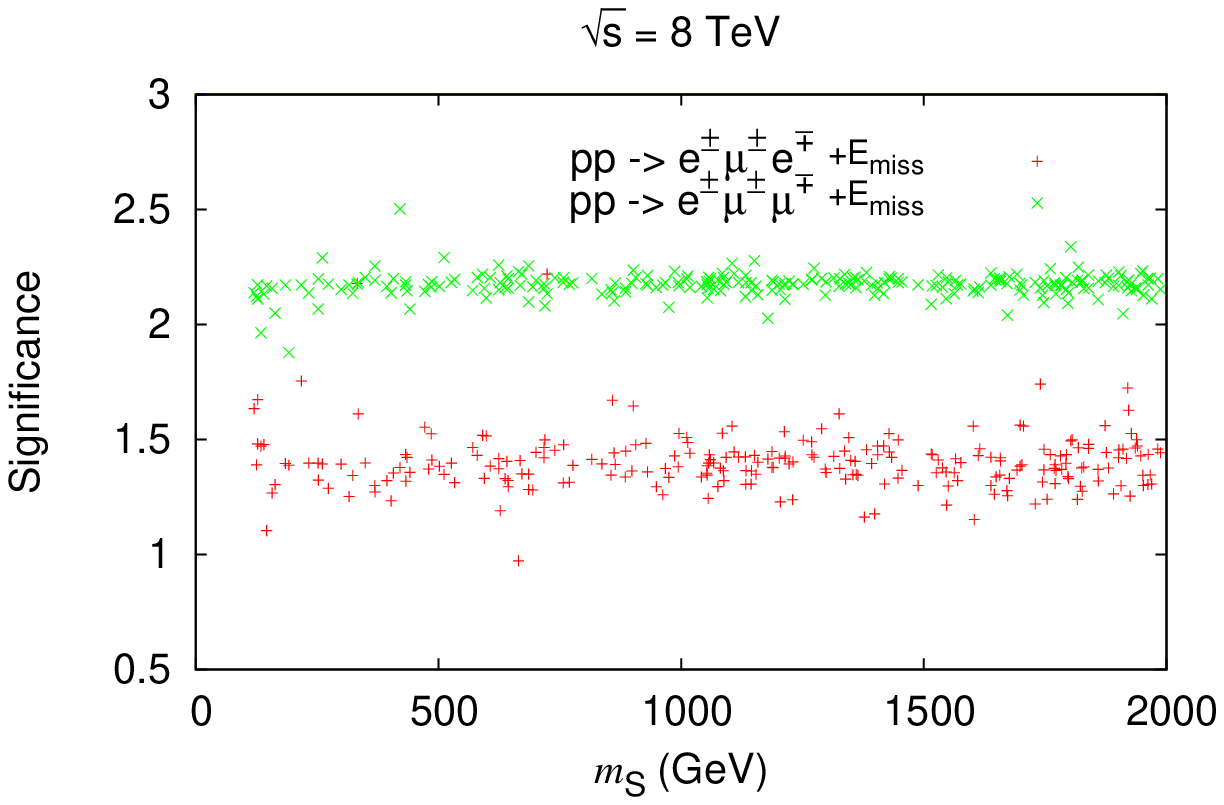}~\includegraphics[width=0.43\textwidth]{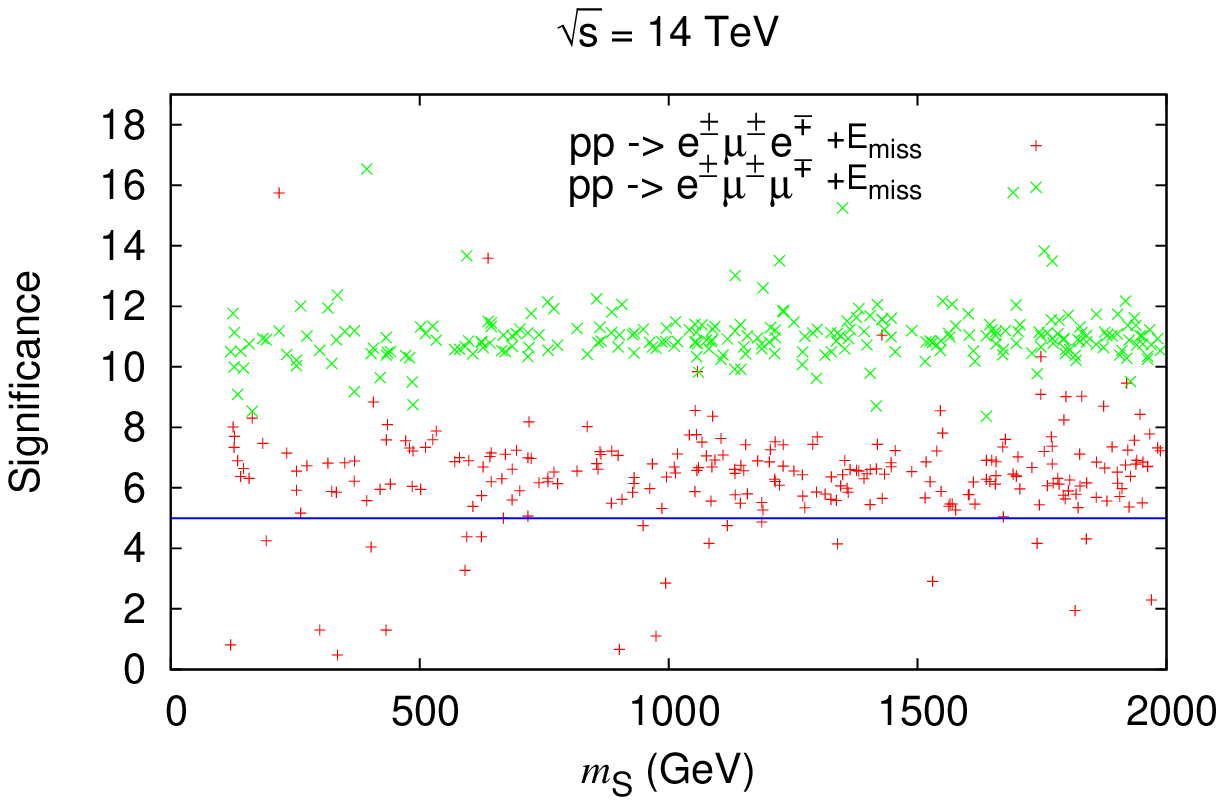}
\end{center}
\caption{Significance for the relevant process $pp\rightarrow\ell^{+}\ell
^{+}\ell^{-}$ + $\slashed{E}_{T}$ at 8 TeV (left) and 14 TeV (right), the horizontal blue line indicate the significance value $S=$ 5.}
\label{fig:sign}
\end{figure}

We look for the event number difference $N_{ex}=N_{M}-N_{B}$, apply the CMS selection criteria used in~\cite{okada} to perform our
analysis, and then compute the significance of each channel for the set of benchmark
points. Figure.~\ref{fig:sign}, shows that it is  possible to find at
least a 1$\sigma$(4$\sigma$) excess in 20.3 $fb^{-1}$ (300
$fb^{-1}$). These results are consistent with searches for new
phenomena since they have not shown any significant deviation from
SM expectations at 8 \textrm{TeV}. Hence, we will select two
benchmark points and apply new cuts for our detailed analysis
thereafter in order to perform the significance signal.

\begin{table}[t]
\caption[]{Two benchmark points selected from the allowed parameter space of
the model.}
\begin{center}
\begin{tabular}
[c]{|c|c|c|c|c|l}\hline Point & $m_{S}$(GeV) & $f_{e\mu}$ & $f_{e\tau}$ &
$f_{\mu\tau}$\\\hline $B_{1}$ & 471.8 &
-$(9.863+i8.774)\times10^{-2}$ & -$(6.354+i2.162)\times 10^{-2}$ &
$(0.78+i1.375)\times10^{-2}$\\\hline $B_{2}$ & 1428.5 &
$(5.646+i549.32)\times10^{-3}$ & -$(2.265+i1.237)\times 10^{-1}$ &
-$(0.41-i3.58)\times10^{-2}$\\\hline
\end{tabular}
\end{center}
\label{tab:points}
\end{table}

\section{Benchmark Analysis and Discussion}

Here, we opt to study the trilepton signal through the two benchmark mass $B_{1}$ and $%
B_{2}$ given in Table.~\ref{tab:points}. We first analyze the trilepton
production with opposite sign same flavor final state at
$\sqrt{s}=8$ and 14 \textrm{TeV}, and then we investigate possibility
of observing the background-free signal $e^{+}\mu^{+}\tau^{-}$.
\newpage

\begin{table}[t]
\caption{Cuts employed for both processes at $\sqrt{s}$ = 8
\textrm{TeV}and $\sqrt{s}$ = 14 \textrm{TeV} respectively.}
\begin{center}
\begin{tabular}
[c]{|l|l|l|l|l|}\hline $e^{+}\mu^{+}e^{-}+\not E_{T}$ &
$e^{+}\mu^{+}e^{-}+\not E_{T}$ & & $e^{+}\mu^{+}\mu^{-}+\not E_{T}$
& $e^{+}\mu^{+}\mu ^{-}+\not E_{T}$ \\\hline
\multicolumn{1}{|c|}{$70<M_{e^{-}e^{+}}<110$} &
\multicolumn{1}{|c|}{$70<M_{e^{-}e^{+}}<110$} &
\multicolumn{1}{|c|}{} &
\multicolumn{1}{|c|}{$80<M_{\mu^{-}\mu^{+}}<100$} &
\multicolumn{1}{|c|}{$80<M_{\mu^{-}\mu^{+}}<110$}\\
\multicolumn{1}{|c|}{$M_{e^{+}\mu^{+}}<200$} & \multicolumn{1}{|c|}{$M_{e^{+}%
\mu^{+}}<230$} & \multicolumn{1}{|c|}{} & \multicolumn{1}{|c|}{$M_{e^{+}%
\mu^{+}}<200$} & \multicolumn{1}{|c|}{$M_{e^{+}\mu^{+}}<230$}\\
\multicolumn{1}{|c|}{$M_{e^{-}\nu}<206$} &
\multicolumn{1}{|c|}{$M_{e^{-}\nu }<220$} & \multicolumn{1}{|c|}{} &
\multicolumn{1}{|c|}{$M_{\mu^{-}\nu}<185$}
& \multicolumn{1}{|c|}{$M_{\mu^{-}\nu}<245$}\\
\multicolumn{1}{|c|}{$10<p_{T}^{\ell}<100$} &
\multicolumn{1}{|c|}{$10<p_{T}^{\ell}<90$} & \multicolumn{1}{|c|}{}
& \multicolumn{1}{|c|}{$10<p_{T}^{\ell
}<100$} & \multicolumn{1}{|c|}{$10<p_{T}^{\ell}<130$}\\
\multicolumn{1}{|c|}{$\left\vert \eta^{\ell}\right\vert <3$} &
\multicolumn{1}{|c|}{$\left\vert \eta^{\ell}\right\vert <3$} &
\multicolumn{1}{|c|}{} & \multicolumn{1}{|c|}{$\left\vert \eta^{\ell
}\right\vert <3$} & \multicolumn{1}{|c|}{$\left\vert
\eta^{\ell}\right\vert
<3$}\\
\multicolumn{1}{|c|}{$\not E_{T}<100$} & \multicolumn{1}{|c|}{$\not
E_{T}<90$} & \multicolumn{1}{|c|}{} & \multicolumn{1}{|c|}{$\not
E_{T}<90$} & \multicolumn{1}{|c|}{$\not E_{T}<120$}\\\hline
\end{tabular}
\end{center}

\label{tab:cuts}
\end{table}

\begin{table}[t]
\caption{The significance corresponding to $%
\mathcal{L}_{int}$ = 20.3 (300) $fb^{-1}$ at 8 and 14 \textrm{TeV}
respectively.}
\label{tab:sign}
\centering
\begin{tabular}
[c]{|c|c||c|c||c|c|}\hline
Process & Benchmark & $N_{20.3}$ & $S_{20.3}$ & $N_{300}$ & $S_{300}$\\\hline
$p,p\rightarrow e^{+}\mu^{+}e^{-}+\slashed{E}_{T}$ & $B_{1}$ & 70.42 & 3.651 &
1689.6 & 17.363\\\cline{2-6}
& $B_{2}$ & 69.69 & 3.618 & 1470 & 15.289\\\hline
$p,p\rightarrow e^{+}\mu^{+}\mu^{-}+\slashed{E}_{T}$ & $B_{1}$ & 71.21 &
3.831 & 2066.7 & 19.210\\\cline{2-6}
& $B_{2}$ & 70.44 & 3.793 & 1974.9 & 18.983\\\hline
\end{tabular}
\end{table}

\subsection{The Processes $ee\mu$\ \& $e\mu\mu$}

To examine the signal discrimination, we focus on the selected
points which are expected to have a favorable cross sections at the
LHC. These points motivate us to the investigation of new cuts on
the relevant observables as shown in Table.~\ref{tab:cuts}, that
would be effective in reducing the backgrounds contribution at
$\sqrt{s} = 8$ and $14$ TeV, where the processes
$pp\rightarrow\ell^{+}\ell^{+}\ell^{-}+\slashed{E}_{T}$ are
mediating by the Feynman diagrams which can be classified as SM and
non-SM diagrams with amplitudes $\mathcal{M}_{SM}$ and
$\mathcal{M}_{S}$, respectively. Therefore,  $N_{ex}=N_{M}-N_{B}$ is directly proportional to
the couplings combination $\left\vert
f_{\alpha\rho}f_{\beta\rho}\right\vert ^{2}$, which means that there
is a direct correlation between the discovery LFV processes and
signals. The corresponding significance computed for each benchmark
point after imposing cuts is shown in Table.~\ref{tab:sign}.
Figure. 4 (left) and (center) exhibits the behavior of the signal
significance which translate the favorable feasibility of detecting
trilepton events through the $\mu^{+}\mu^{-}$ signature.

\begin{figure}[h]
\begin{center}
\includegraphics[width=0.33\textwidth]{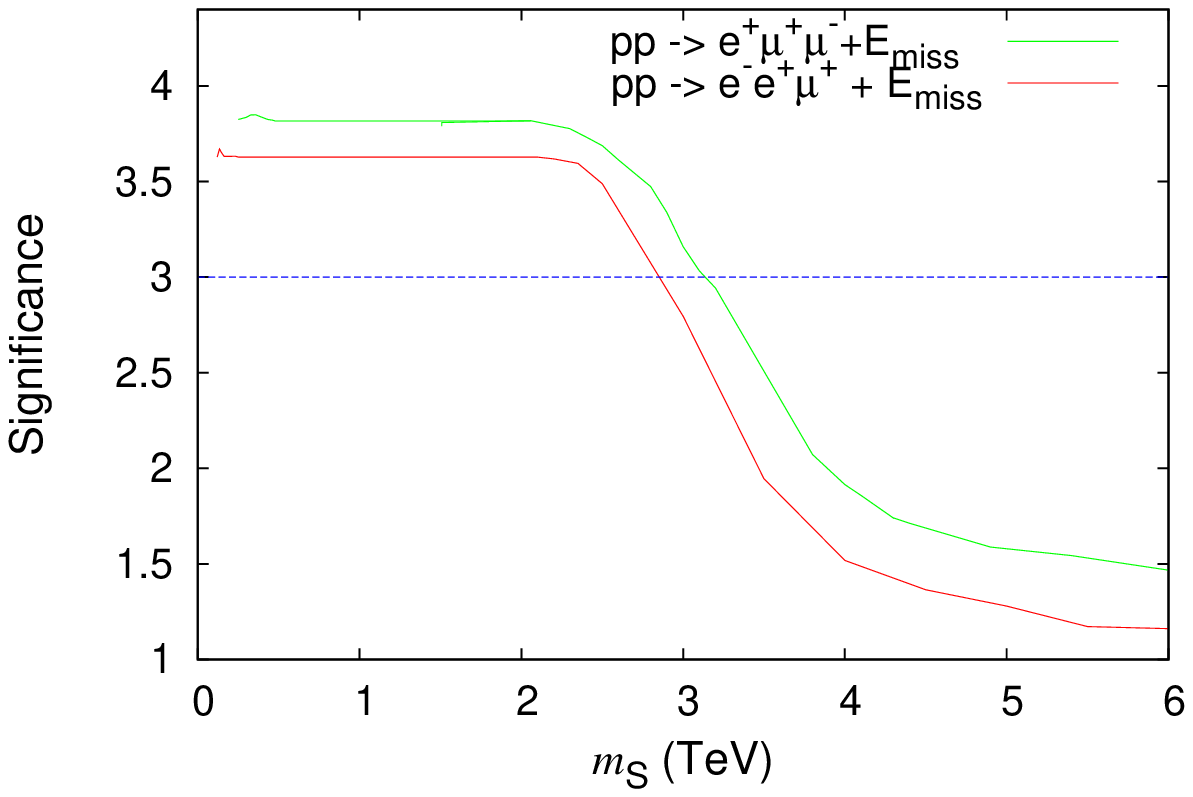}~\includegraphics[width=0.33\textwidth]{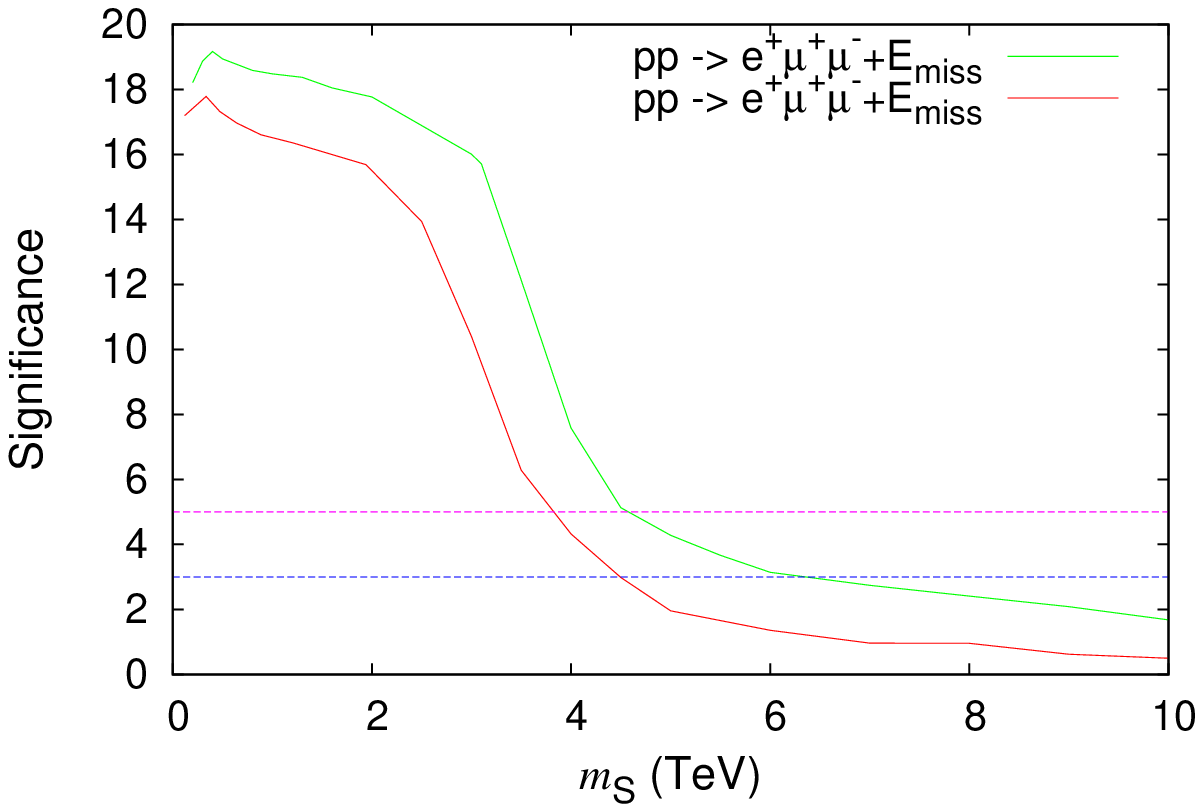}~\includegraphics[width=0.33\textwidth]{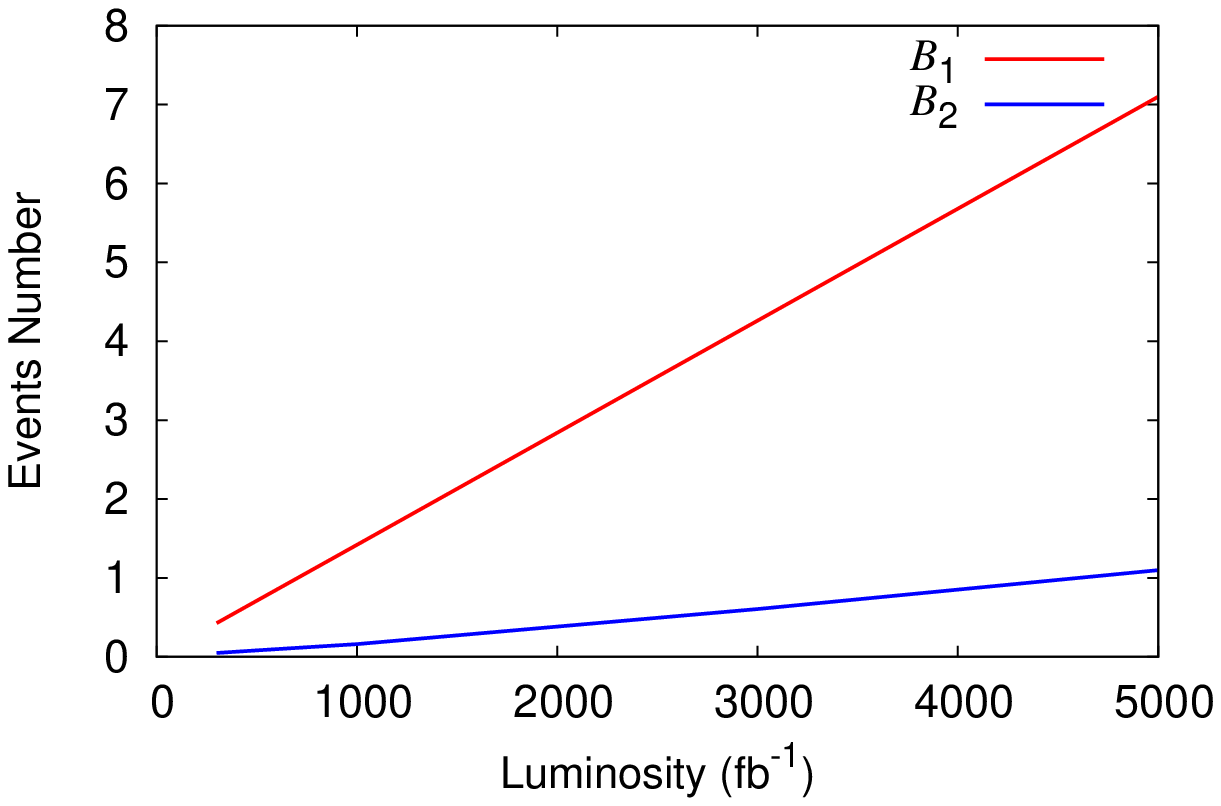}
\caption{Significance for the process $pp \rightarrow\ell^{+}\ell
^{+}\ell^{-} + \slashed{E}_{T}$ at $\sqrt{s}$ = 8 \textrm{TeV}
(left) and $\sqrt{s}$ = 14 \textrm{TeV} (center) within new cuts, the
dashed horizontal blue (pink) line indicate the significance value $S=$ 3 ($%
S=$ 5) respectively. In
(right) events number for the background-free process $pp\rightarrow
e^{+}\mu^{+}\tau^{-}+\slashed{E}_{T}$ at $\sqrt{s}$= 14
\textrm{TeV}.} \label{s}
\end{center}
\end{figure}

\subsection{LFV\ Background Free Channel}

To further our investigation, we extend our earlier analysis in the
perspective of optimize the detection of this signature in colliders
for both benchmark points through the background free process
$e^{+}\mu^{+}\tau^{-}$. However, observing such process requires huge luminosity and the resulting number of events is very low (less than 3 events for $1000~\text{fb}^{-1}$ lumunoisity).

\subsection*{Acknowledgments}

D. Cherigui  would like to thank the organizers of the Moriond Conference   for the
financial support.

\end{document}